\documentclass[twocolumn,showpacs,amsmath,amssymb,superscriptaddress,aps,prl]{revtex4-1}
\usepackage{graphicx}
\usepackage{dcolumn}
\usepackage{bm}
\usepackage{epsf}
\usepackage{amsmath}

\usepackage{amssymb}
\usepackage{lipsum}
\usepackage{blindtext}

\newcommand{\be}{\begin{equation}}
\newcommand{\ee}{\end{equation}}

\newcommand{\ket}[1]{|#1\rangle}

\newcommand{\tr}[1]{\mathrm{tr}\left\{#1\right\}}

\newcommand{\bla}{bla\\bla\\bla\\bla\\bla}

\usepackage[usenames,dvipsnames]{color}
\usepackage{ulem}
\usepackage{soul}
\setstcolor{red}
\sethlcolor{yellow}

\begin{document}

\title{Thermodynamics of weakly measured quantum systems}
\author{Jose Joaquin Alonso}
\affiliation{Department of Physics, Friedrich-Alexander-Universit\"at Erlangen-N\"urnberg, D-91058 Erlangen, Germany}
\author{Eric Lutz}
\affiliation{Department of Physics, Friedrich-Alexander-Universit\"at Erlangen-N\"urnberg, D-91058 Erlangen, Germany}
\author{Alessandro Romito}
\affiliation{Dahlem Center for Complex Quantum Systems, FU Berlin, D-14195 Berlin, Germany}

\date{\today}

\begin{abstract}
We consider  continuously monitored quantum systems and introduce definitions of work and heat along individual quantum trajectories that are valid for coherent superposition of energy eigenstates. We use these quantities to extend the first and second laws of stochastic thermodynamics to the quantum domain. We illustrate our results with the case of a weakly measured driven two-level system and show how to distinguish between quantum work and heat contributions. We finally employ quantum feedback control to suppress detector backaction and determine the work statistics.

\end{abstract}
\pacs{03.65.Yz, 05.30.-d}
\maketitle

Thermodynamics is, at its heart, a theory of work and heat. The first law is based on the realization that both quantities are two forms of energy and that their sum is conserved. At the same time, the fact that  entropy, defined as the ratio of  reversible heat and temperature, can only increase in an isolated system is an expression of the second law \cite{pip66, cal85}. In classical thermodynamics, work is defined as the change of internal energy in an isolated system, $W=\Delta U$, while {heat is introduced as the difference},  $Q=\Delta U -W$, in a nonisolated system. Thermal isolation is thus crucial to distinguish $W$ from $Q$. In the last decades, stochastic thermodynamics has successfully extended the definitions of  work and heat to the level of single trajectories of microscopic systems \cite{sei12}. In this regime, thermal fluctuations are no longer negligible  and the laws of thermodynamics have  to be adapted to fully include them. The second law has, for instance, been generalized in the form of fluctuation theorems that quantify the occurrence of negative entropy production \cite{jar11}. A particular example is the Jarzynski equality, $\langle \exp(-\beta W)\rangle = \exp(-\beta \Delta F)$, that allows the determination of equilibrium free energy differences $\Delta F$ from the nonequilibrium work statistics in systems at initial inverse temperature $\beta$ \cite{jar97}. The laws of stochastic thermodynamics have been verified in a large number of different experiments, see  Refs.~\cite{lip02,bli06} and the review \cite{cil13}.

The current challenge  is to extend the principles of thermodynamics to include quantum effects which are expected to dominate at smaller scales and colder temperatures.  Some of the unsolved key issues concern the correct definition of quantum work and heat, means to distinguish between the two quantities owing to the blurring effect of quantum fluctuations, and   the proper clarification of the role of quantum coherence. A variety of approaches have been suggested  to tackle these   problems  \cite{tal07,hub08,esp09,cam11,hor12,hek13,dor13,maz13,abe13,hal14,ron14,gal15},
and quantum work statistics has been measured in isolated systems in two pioneering experiments using NMR \cite{bat14} and trapped ions \cite{an14}. A new approach  may emerge from the possibility of weakly monitoring quantum systems. Recently,  individual quantum trajectories of a superconducting qubit in a microwave cavity have been observed using weak measurements \cite{mur13,web14}.
These measurements only slightly disturb quantum systems owing to the weak coupling to the measuring device \cite{cle10}. They hence allow to gain information about  states without projecting them into eigenstates. They have been successfully employed to explain quantum paradoxes \cite{aha05}, detect and amplify weak signals \cite{hos08,dix09}, determine a quantum virtual state \cite{romito14}, as well as directly measure a wave function \cite{lun11}. Motivated by the two experiments  \cite{mur13,web14}, we here investigate the first and second law for continuously monitored quantum systems and aim at developing a quantum stochastic thermodynamics based on quantum trajectories. Such an extension faces several technical difficulties. First, since a weakly measured system can be in a coherent superposition of energy eigenstates, energy is not a well-defined concept along a single quantum trajectory. Furthermore, even in the absence of an external environment, a continuously monitored quantum system is not isolated and the detector backaction, albeit small, will perturb its dynamics \cite{cle10}. As a result, its time evolution  will be nonunitary  and energy, in the form of heat, will be exchanged with the detector.

 In the following, we introduce  suitable and consistent definitions of work and heat contributions to the quantum stochastic evolution of a  weakly measured system that is externally driven. 
 We  use these definitions to determine the distributions of  quantum work and heat for a  two-level system, and demonstrate the general validity of the Jarzynski equality, hence of the second law. We finally use the tools of quantum feedback control \cite{wis10} to suppress detector backaction and thus effectively achieve thermal isolation of the system. This provides  a practical  scheme to experimentally test our definitions of work and heat along individual quantum trajectories.


\paragraph{Quantum work and heat.} We consider a  system with time-dependent Hamiltonian $H_t$ that is initially in a thermal  state at inverse temperature $\beta$, $\rho_0= \exp(-\beta H_0)/Z_0$, where $Z_0$ is the partition function. The system  is driven by an external parameter $\lambda_t$ during a time  $\tau$. At the ensemble level, quantum work and heat are introduced by considering an infinitesimal variation of the mean energy, $U = \langle H\rangle = \tr{\rho_t H_t}$ \cite{rei65,per93}:
\be
\label{1}
dU = \tr{\rho_t dH_t} + \tr{d\rho_t H_t} = \delta W+ \delta Q.
\ee
Heat is further related to entropy $S=-k \text{tr}\{\rho_t \ln \rho_t\}$ via $\delta Q = T dS$ \cite{rei65,per93}. For an isolated system with unitary dynamics heat vanishes, since $dS=0$, and therefore $dU = \delta W$ in agreement with classical thermodynamics \cite{pip66, cal85}. Heat therefore appears to be fundamentally associated with the nonunitary part of the dynamics.

At the level of individual realizations, energy is a stochastic quantity owing to  thermal and quantum fluctuations. The distribution $p(u$) of the total energy change $u$ may be determined by performing projective measurements  $\Pi_n$ and $\Pi_m$, with  outcomes $E^0_n$ and $E^\tau_m$, at the beginning and at the end of the driving protocol \cite{tal07,kaf12},  
\begin{eqnarray}
\label{prob_du}
p(u)=\sum_{m,n}P^\tau_{m,n}P^0_n\delta(u-\Delta E_{m,n}).
\end{eqnarray}
Here $P^0_n= \tr{\Pi_n\rho_0}$ denotes the probability  of the   eigenvalue $E_n^0$, $P^\tau_{m,n}= \tr{\Pi_m \rho_{n,\tau}}$ the transition probability from state $n$ to $m$,  with  $\rho_{n,\tau}$ the time evolved projected density operator $\rho_{n,0}=\Pi_n \rho_0\Pi_n/P^0_n$, and $\Delta E_{m,n} = E^\tau_m- E^0_n$ the energy difference. For unitary dynamics, $p(u)$ reduces to the work distribution $p(W)$, but, in general,  Eq.~\eqref{prob_du} does not allow  to distinguish  work from heat. In the following, we generalize  Eq.~\eqref{prob_du} and identify work and heat for a weakly measured  system. 

A  quantum system continuously monitored by a quantum limited detector may be assigned, for each individual  trajectory, a conditional density operator $\tilde \rho_t$ that reduces to the usual density operator $\rho_t$ when averaged over all the trajectories, $\rho_t= \langle\langle \tilde \rho_t\rangle\rangle$ \cite{wis10,jac14}. The evolution of $\tilde \rho_t$ is commonly described  by a stochastic master equation that contains  a random parameter $\xi(t)$ that accounts for the detector shot noise, see Eqs.~\eqref{rho_evol1}-\eqref{rho_evol2} below for an example. An important observation is that such master equation has  a unitary component, corresponding to the dynamics generated by the system's Hamiltonian, and a nonunitary part that stems from the continuous coupling to the detector.  For an infinitesimal time step, these two contributions  are additive and may be written as,
\begin{align}
\label{primo-principio}
d \tilde{\rho}_t= \delta \mathbb{W}[\tilde \rho_t] dt+ \delta \mathbb{Q}[\tilde \rho_t]dt,
\end{align}
where $\delta \mathbb{W}[\tilde \rho_t] $ and $\delta \mathbb{Q}[\tilde \rho_t] $ are  operators associated with the respective unitary and nonunitary parts of the dynamics. We identify them as corresponding to work and heat at the level of an infinitesimal quantum trajectory.  This separation cannot be directly extended to the entire (time integrated) trajectory, since the stochastic master equation is generally a nonlinear function of  the operator $\tilde \rho_t$. However, when averaged over quantum fluctuations, Eq.~\eqref{primo-principio} allows to extend the first law \eqref{1} to single realizations of the stochastic measurement outcome, 
\begin{align}
\label{primo-principio-stocastico}
d \tilde U_t=& \tr{H_t(\tilde\rho_{t-dt}+d\tilde\rho_t)}-\tr{H_{t-dt}\tilde\rho_{t-dt}}\nonumber\\
=& \tr{ \tilde\rho_{t-dt}dH_t}+\tr{H_{t}\delta\mathbb{W}dt } + \tr{H_{t}\delta\mathbb{Q}dt }\nonumber\\
=&\delta \tilde W_t + \delta \tilde Q_t,
\end{align} 
where in the second line $dH_t=H_t-H_{t-dt}$ and the middle term  $\text{tr}\{H_t\delta\mathbb{W}dt\}=0$ since $\delta\mathbb{W}$ is unitary \cite{unitary}. In the last line, $\delta\tilde W_t=\tr{\tilde\rho_{t-dt}dH_t}$ and $\delta\tilde Q_t=\tr{H_{t}\delta\mathbb{Q}dt }$, indicating that work is related to a change of the Hamiltonian, as expected, and heat to the nonunitary $\delta\mathbb{Q}$.  Equation \eqref{primo-principio-stocastico} is a direct extension of stochastic thermodynamics to the quantum domain. The  first law \eqref{1} is recovered when Eq.~\eqref{primo-principio} is averaged over both stochastic and quantum fluctuations.
The integrated work and heat contributions to the changes in transition probabilities, $d\tilde P_{m,n}=\tilde P^\tau_{m,n}-\tilde P^0_{m,n}$, with $P^0_{m,n}=\delta_{n,m}$ the initial transition probability, can be further obtained from Eq.~\eqref{primo-principio} by carefully adding all the different terms (see Ref.~\cite{suppl}). We find, for each individual quantum trajectory, 
\begin{align}
d\tilde P_{m,n} =\delta\tilde P^W_{m,n}+\delta\tilde  P^Q_{m,n},\label{main}
\end{align}
with the two quantities,
\begin{eqnarray}
\delta\tilde P^W_{m,n}&=&  \tr{ \Pi_m \int_0^\tau dt \, \delta\mathbb{W} [\tilde \rho_t]}, \label{5}
\\
\delta\tilde P^Q_{m,n}&=& \tr{\Pi_m \int_0^\tau dt\, \delta \mathbb{Q}[\tilde \rho_t]} \label{6}.
\end{eqnarray}
These expressions depend explicitly on the quantum trajectory $\tilde \rho_t$ which  we  stress  by using the notation  $\delta$ instead of $d$. They provide an unambiguous way to distinguish between work and heat  at the level of a single trajectory. Remarkably, they are valid even if the system remains in a coherent superposition of energy eigenstates, that is, when its energy  is ill-defined. 
 Equation \eqref{main} holds for the trajectory averaged quantities $dP_{m,n}= \delta P^W_{m,n}+\delta P^Q_{m,n}$, with $\delta P^\alpha_{m,n}=\langle\langle \delta \tilde P^\alpha_{m,n}\rangle\rangle$, $\alpha = W, Q$. This averaged distinction between work and heat contributions to transition probabilities requires access to single trajectories, thus  cannot be established directly at the ensemble level.

The second law in the form of the Jarzynski equality, $\int dW p(W) \exp(-\beta W) = \exp(-\beta \Delta F)$, immediately follows from Eq.~\eqref{prob_du} for an isolated system \cite{tal07}. However, the equality is not satisfied for an open system with nonunitary dynamics owing to the heat term \cite{kaf12}. The second law may be restored by replacing $P^\tau_{m,n}$ by $P^W_{m,n}=P^0_{m,n} + \delta P^W_{m,n}$, that is, by setting $\delta P^Q_{m,n}$ to zero at each time step, see Fig.~3.  We  next show how   quantum work and heat may be identified theoretically, by numerically analyzing a weakly measured   two-level system, and experimentally, by means of quantum feedback control.

\begin{figure}[t]
\includegraphics[width=0.48\textwidth]{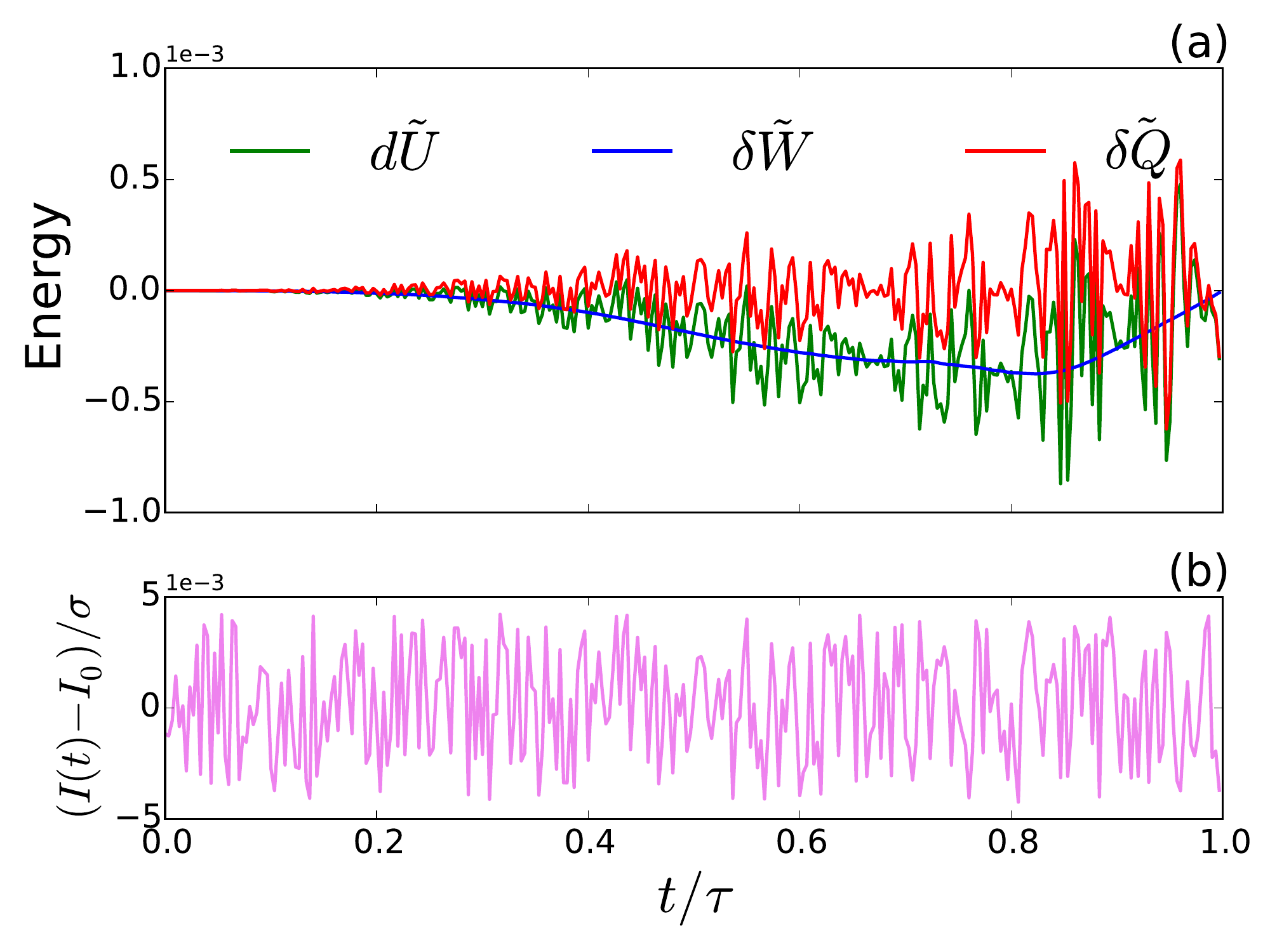}
\caption{\label{trajectory} (color online) First law   for a weakly measured qubit. a) Infinitesimal change of work, heat and  energy along a single quantum trajectory $\tilde \rho_t$; for each realization  $d\tilde U_t =  \delta \tilde W_t + \delta \tilde Q_t$, Eq.~(4). b) Corresponding signal $I(t)$ in the  detector. Parameters are $S_0/\Delta I^2=2.5\cdot 10^5dt$, $\hbar/g=1.6\cdot10^2dt$, $\hbar/\epsilon=10^3dt$, $\nu =8$  and  $\tau=3\cdot 10^3dt$ (see main text).} 
\end{figure}


\paragraph{Application to a monitored qubit.}
In order to illustrate our approach, we consider a driven two-level system $\textit{S}$  with Hamiltonian, $H_{t}=\epsilon\sigma_z+\lambda_t\sigma_x$, where $\lambda_t$ is the external driving and $\sigma_i$  the usual Pauli matrices. The system is continuously coupled to a quantum limited detector \textit{D} via the interaction Hamiltonian $H_{\textit{I}}$:
\begin{align}
H=H_t+H_{\textit{D}}+\sigma_z H_{\textit{I}},\label {modello}
\end{align}
where, without loss of generality, we identify $\sigma_z$ as the system's observable that is monitored by the detector. The effect of the detector is fully characterized by the averaged signals, ($I_1$, $I_2$), and  Gaussian noises, ($S_1$,  $S_2$), measured when the qubit is in the two eigenstates, ($\ket{1}$, $\ket{2}$), of the measured observable. 
We assume to be in the weak measurement regime, i.e. at time scales smaller than the measurement time $\tau_M=(S_1+S_2)/(I_1-I_2)^2$.
For concreteness and simplicity, we will interpret the qubit in Hamiltonian \eqref{modello} as describing a double quantum dot sharing a single electron and interacting with a quantum point contact (QPC), but it can also be applied to a qubit coupled to a microwave resonator \cite{kor11} in a circuit QED set-up as in the experiments \cite{mur13,web14}.
 We accordingly identify the configurations where the electron occupies only one dot by $\langle \sigma_z \rangle=\pm 1$. Coherent superpositions of the two are possible. The detector monitoring the occupation of the dots is a voltage biased QPC with Hamiltonian \cite{gurvitz97, korotkov99,korotkov01, korotkov02} 
$H_{\textit{D}}=\sum_{l}E_l a^{\dagger}_la_l+\sum_{r}E_ra^{\dagger}_ra_r+
\sum_{l,r} \Omega (a^{\dagger}_ra_l+a^{\dagger}_la_r)$ and the interaction term reads $H_I=\sum_{l,r}\delta \Omega/2 (a^{\dagger}_ra_l+a^{\dagger}_la_r)$.
The signal in the detector is the current $I(t)$ across the QPC, with averages  $I_{1(2)}=2\pi\Omega_{1(2)}^2\rho_l\rho_re^2V/\hbar = e^2 T_{1(2)} V/h$  and  noises $S_{1(2)} = e(1-T_{1(2)}) I_{1(2)}$. Here $\rho_{l,r}$ are the densities of states in the left and right electrodes, and $T_{1(2)}$ are the dimensionless transmission probabilities across the QPC. 

\begin{figure}[t]
\includegraphics[width=0.42\textwidth]{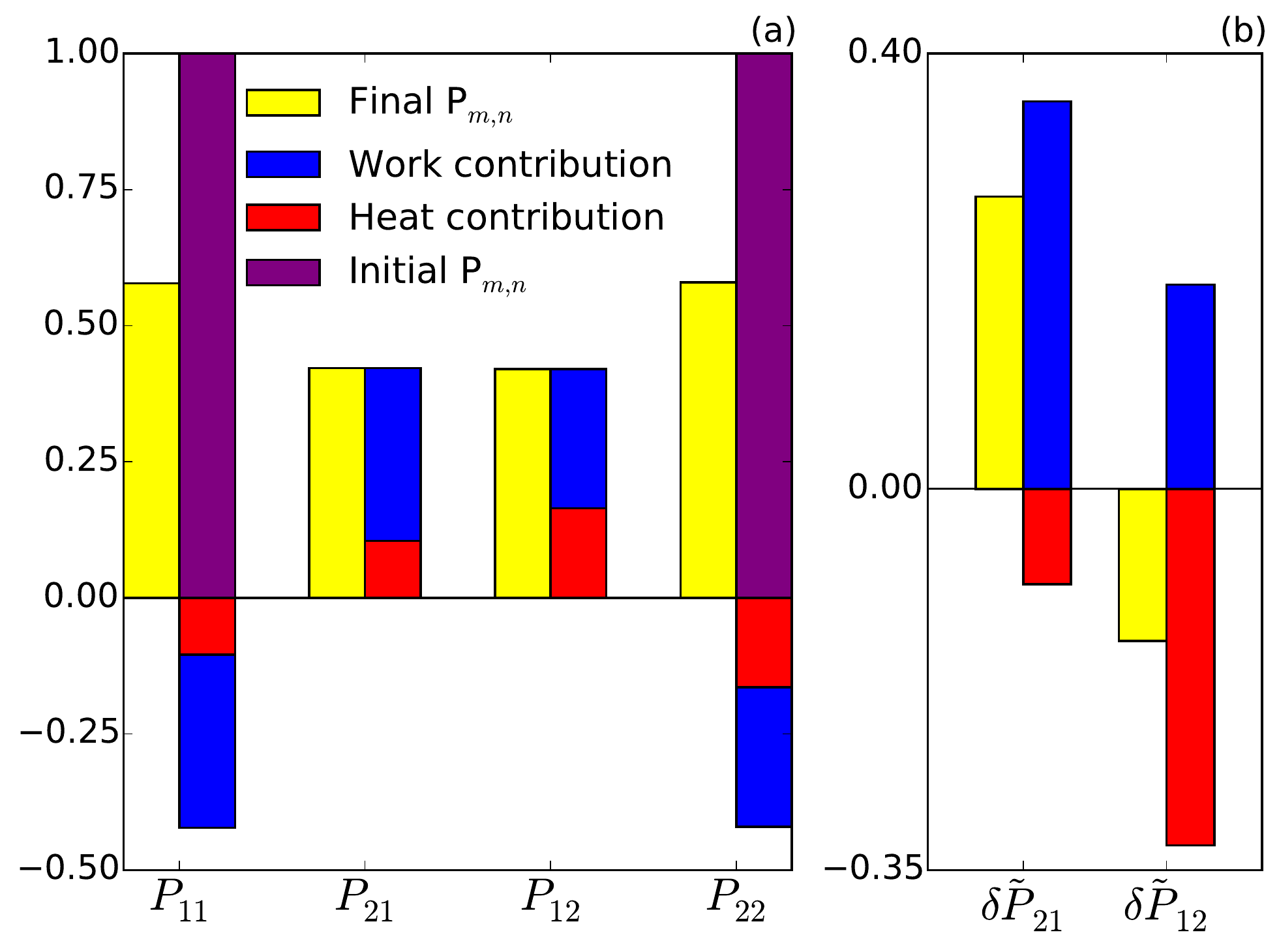}
\caption{\label{distrib_meas} (color online) a) Averaged final transition probabilities $P^\tau_{m,n}$ (yellow) for a continuously monitored qubit  with their  work and heat contributions, $\delta P_{m,n}^W$ (blue) and $\delta P_{m,n}^Q$ (red), and the initial transition probability, $P^0_{m,n}=\delta_{mn}$ (purple). The first law like equation $dP_{m,n}= \delta P^W_{m,n}+\delta P^Q_{m,n}$ is verified. b) Work and heat contributions, $\delta \tilde P_{m,n}^W$  and $\delta \tilde P_{m,n}^Q$  at the single trajectory level. Same parameters as in Fig.~\ref{trajectory}.}
\end{figure}

Under the assumption of a weakly coupled detector, the detector signal is a random variable, and the evolution of the system depends on the specific realization of the stochastic process. This is captured by a well-established Bayesian formalism~\cite{korotkov01, korotkov02} which describes the evolution of the system conditional to the detector's outcome in terms of a nonlinear stochastic differential equation for the system's density matrix $\tilde \rho (t)$. In the Ito formulation, we have \cite{korotkov01, korotkov02},
\begin{align}
\label{rho_evol1}
\dot{{\tilde \rho}}_{11}=&- 2\frac{\lambda(t)}{\hbar}\operatorname{Im}\left({\tilde \rho}_{12}\right)+{\tilde \rho}_{11}(1-{\tilde \rho}_{11})\frac{2\Delta I}{S_0}\xi(t),\\
\label{rho_evol2}
\dot{{\tilde \rho}}_{12}=&2i\frac{\epsilon}{\hbar}{\tilde \rho}_{12}-i\frac{\lambda(t)}{\hbar}\left(1-2 {\tilde \rho}_{11}\right)-{\tilde \rho}_{12}\frac{(\Delta I)^2}{4S_0}\nonumber \\
& +(1-2{\tilde \rho}_{11}){\tilde \rho}_{12}\frac{\Delta I}{S_0}\xi(t),
\end{align}
where  $\Delta I=I_2-I_1$ and $\xi(t)$  is the white noise of the detector's signal with $\langle \xi(t)\rangle =0$, $\langle \xi(t) \xi(t')\rangle =\sigma^2 \delta (t-t')$  and $\sigma=\sqrt{S_0/2}$. The detector current $I(t)$  is further,
\begin{eqnarray}
\label{current}
I(t)= I_0+\frac{\Delta I}{2}(2\tilde{\rho}_{11}-1)+\xi(t).
\end{eqnarray}
 For each realization of the measurement outcome,  Eqs.~\eqref{rho_evol1} and \eqref{rho_evol2}  allow to identify the unitary and nonunitary contributions to the time evolution, since the nonunitary part is proportional to $\Delta I$. We  rewrite Eq.~\eqref{primo-principio} as  $d \tilde \rho_t= \delta \mathbb{W}[\tilde \rho_t] dt+ (\Delta I/S_0 ) \delta \mathbb{M}[\tilde \rho_t]dt$, and identify $\delta \mathbb{W}$ with the work done by the  driving  $\lambda_t$ along an infinitesimal trajectory and $\delta \mathbb{Q}=(\Delta I/S_0) \delta \mathbb{M}$ as the heat  associated with the detector backaction of the detector. Due to the nonlinearity of the stochastic master equation, we can only determine the  distributions of work and heat numerically. We specify the driving as $\lambda_t=g(1/\cosh(\nu(1- t/\tau))$, where $\tau$ is the duration of the experiment, and reformulate equations \eqref{rho_evol1} and \eqref{rho_evol2} in the Stratonovich form~\cite{korotkov01, korotkov02}. We solve them  numerically by the Monte-Carlo method  for an ensemble of 300 realizations of the random signal $I(t)$ in the interval $t/\tau \in[0,1]$ using a  time step  $dt=0.01$. The  results  for work, heat and energy  along a given quantum trajectory, Eq.~\eqref{primo-principio-stocastico}, are shown in Fig.~\ref{trajectory}, while  those for the work and heat  contributions to the transition amplitudes, Eqs.~\eqref{5}-\eqref{6}, are presented in  Fig.~\ref{distrib_meas} (see Ref.~\cite{suppl} for details).

Figure~\ref{trajectory}a) demonstrates  the reconstruction of  quantum averaged work and heat changes, $\delta \tilde W_t$ and  $\delta \tilde Q_t$,  along a single quantum trajectory, based on the definitions given in Eq.~(4). The corresponding  signal $I(t)$  in the detector is displayed in Fig.~\ref{trajectory}b). Contrary to the case of an isolated system for which  $d \tilde U=\delta \tilde W$, the heat contribution $\delta \tilde Q_t$ is here clearly visible. Equation\eqref{primo-principio-stocastico} holds for each individual realization and thus extends the first law of stochastic thermodynamics to the quantum regime. Figure \ref{distrib_meas}b) shows the unambiguous distinction of the work and heat contributions, $\delta \tilde P^W_{m,n}$ and $\delta \tilde  P^Q_{m,n}$, evaluated via Eqs.~\eqref{5}-\eqref{6}, to the final transition probability $P^\tau_{m,n}$. We stress that, although $P^\tau_{m,n}$ is always positive, as a proper probability should be, the work and heat contributions need not be: the  probability to go from state $n$ to $m$ at time $\tau$ can, for instance, be smaller than the initial transition probability \cite{rem}. Note that a quantity, $dP_{m,n}^\alpha=P_{m,n}^{\alpha,\tau}-P^{\alpha,0}_{m,n}, \alpha=W, Q$, that only depends on initial and final times,  cannot be defined, reflecting the fact that there are no heat or work operators.

\begin{figure}[t]
\includegraphics[width=0.47\textwidth]{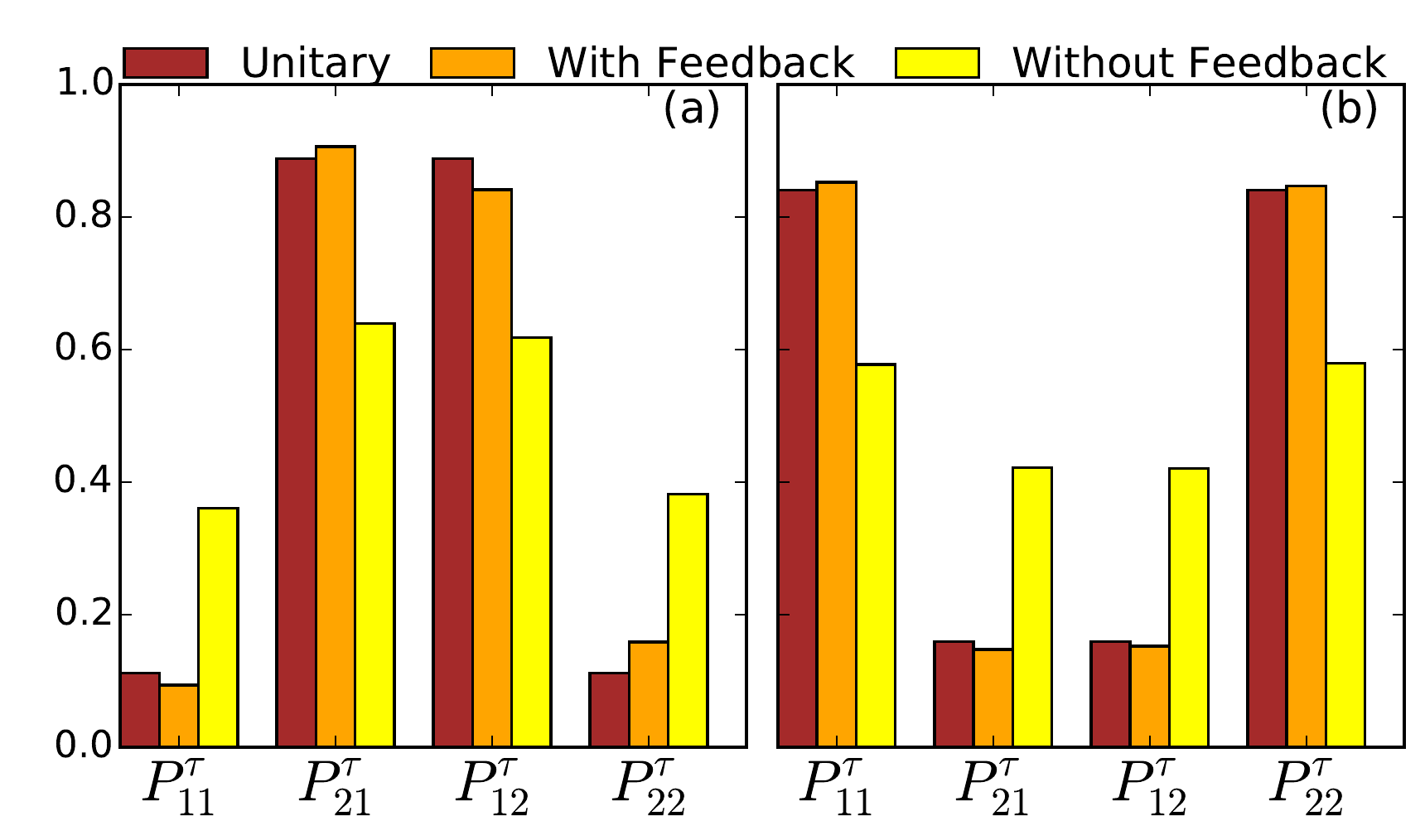}
\caption{\label{distrib} (color online) Transition probabilities $P^\tau_{mn}$ for the weakly measured qubit with (orange) and without (yelllow)  feedback control for a)   $\tau_1=1.4\cdot 10^3dt $  and b) $\tau_2= 2.5\cdot 	10^3 dt$. The feedback strength is $f=3$. The isolated, unitary, case  (red) is shown as a reference. The feedback loop effectively suppresses the detector backaction and the associated heat exchange,  achieving thermal isolation.}
\end{figure}


\paragraph{Quantum feedback control.} 
In classical thermodynamics work is associated with the variation of the internal energy of the isolated system \cite{pip66, cal85}. After having shown above how heat can be theoretically identified, we next take advantage of a feedback loop protocol to suppress the detector backaction \cite{wis10}, offering a scheme to reach isolation experimentally.  Quantum feedback has recently been demonstrated experimentally for a superconducting qubit \cite{mur12}.
Specifically we control the amplitude, $g$, of the system's driving depending on the continuos detector outcome, i.e. $g \to g_t \equiv (1-f \Delta \varphi_t) g$,
where $f$ is the feedback strength and  $\Delta \varphi_t$ the phase difference between the actual vector (with backaction) and desired vector (without backaction)  in the Bloch sphere of the qubit (see Refs.~\cite{korotkov01,suppl} for details). This allows  to operationally counter the effects induced by the continuous monitoring. From a thermodynamic point of view, the feedback adds an extra amount of work that exactly cancels the heat contribution to  the transition probabilities.

Figure \ref{distrib} shows the numerically simulated  final transition probabilities $P^\tau_{m,n}$ for the weakly measured qubit with (orange) and without (yellow) quantum  feedback for two driving times. 
We observe in both cases that the feedback process effectively suppresses the heat contributions (identified in Fig.~\ref{distrib_meas}) and that the transition probabilities agree with those of the isolated system with unitary dynamics (brown). Quantum feedback control thus appears as a powerful tool to determine the statistics of the work done by the external driving in a continuously monitored  system. The heat statistics can be further easily  obtained by measuring the undriven system, that is, when no work is performed and $d\tilde U_t = \delta \tilde Q_t$.

The above findings can be directly used to verify the quantum Jarzynski equality for the driven qubit. Since any measurement induced heating is prevented by the feedback, only the initial inverse temperature $\beta$ of the system matters. Determining the quantum work statistics via Eq.~(2), we find $\Delta F_1 = -0.488$ and  $\Delta F_2= -0.496$  with feedback control for $\tau_1 =  1.4\cdot 10^3dt$  and $\tau_2= 2.5\cdot 10^3dt $  and $\Delta F = -0.495$ in the unitary case (for  $\beta =10$).  The excellent agreement demonstrates the correctness of the  definitions of work and heat, and confirms the second law for a weakly measured quantum system.
%


\paragraph{Conclusions.} We have  extended the laws of stochastic thermodynamics along individual quantum trajectories of a weakly measured system. We have shown how to distinguish work and heat contributions to both the energy changes and the transition probabilities. We have further demonstrated the usefulness of our approach with the analysis of a driven qubit and introduced methods to identify work from heat numerically as well as experimentally with the help of quantum feedback control.

\paragraph{Acknowledgments} 
This work was partially supported by DFG under Grants No. RO 4710/1-1 and LU 1382/4-1, the EU Collaborative Project TherMiQ (Grant Agreement 618074) and the COST Action MP1209.

\paragraph{Note added.} While completing this manuscript, we became aware of  a preprint by Elouard {\it et al.} \cite{Elouard15} that also discusses quantum stochastic thermodynamics.


\begin{thebibliography}{99}
\bibitem{rem} An insightful discussion of the role and interpretation of negative probabilities in physics has been provided by Feynman \cite{fey87}.
\end{thebibliography}
\end{document}